\documentclass[journal]{IEEEtran}

\usepackage{cite}
\usepackage{graphicx}
\usepackage{amsmath}
\usepackage{amsfonts}
\usepackage{algorithmic}
\usepackage{textcomp}
\usepackage{xcolor}

\begin{document}

\title{Multiscale micromagnetic / atomistic modeling of heat assisted magnetic recording}

\author{
    \IEEEauthorblockN{
        Mohammed Gija\IEEEauthorrefmark{1}
        Alexey Dobrynin\IEEEauthorrefmark{2},
        Kevin McNeill\IEEEauthorrefmark{2}, 
        Mark Gubbins\IEEEauthorrefmark{2},
        Tim Mercer\IEEEauthorrefmark{1},
        Philip Bissell\IEEEauthorrefmark{1},
        Serban Lepadatu\IEEEauthorrefmark{1}
    }
    \\

     \IEEEauthorblockA{\IEEEauthorrefmark{1} Jeremiah Horrocks Institute, University of Central
Lancashire, Preston PR1 2HE, UK}

    \IEEEauthorblockA{\IEEEauthorrefmark{2}Seagate Technology, 1 Disc Drive, Derry BT48 0BF, Northern Ireland, UK}
\\

}

\maketitle

\begin{abstract}
Heat-assisted magnetic recording (HAMR) is a recent advancement in magnetic recording, allowing to significantly increase the areal density capability (ADC) of hard disk drives (HDDs) compared to the perpendicular magnetic recording (PMR) technology. This is enabled by high anisotropy FePt media, which needs to be heated through its Curie temperature ($T_C$) to facilitate magnetization reversal by an electromagnetic write pole. HAMR micromagnetic modeling is therefore challenging, as it needs to be performed in proximity to and above $T_C$, where a ferromagnet has no spontaneous magnetization. An atomistic model is an optimal solution here, as it doesn't require any parameter renormalization or non-physical assumptions for modeling at any temperature. However, a full track atomistic recording model is extremely computationally expensive. Here we demonstrate a true multiscale HAMR modeling approach, combining atomistic spin dynamics modeling for high temperature regions and micromagnetic modeling for lower temperature regions, in a moving simulation window embedded within a long magnetic track. The advantages of this approach include natural emergence of $T_C$ and anisotropy distributions of FePt grains.  Efficient GPU optimization of the code provides very fast running times, with a 60~nm wide track of twenty-five 20~nm - long bits being recorded in several hours on a single GPU. The effects of realistic FePt L$_{10}$ vs simple cubic crystal structure is discussed, with the latter providing further running time gains while keeping the advantages of the multiscale approach. 
\end{abstract}


\section{Introduction}
\IEEEPARstart{H}{ard} disc drives (HDDs) remain the key data storage technology due to their long-term reliability and low cost per GB. The International Data Corporation Global DataSphere Forecast estimated that by 2028 the worldwide data created, captured and duplicated will increase to 394~ZB~\cite{HDD}. Therefore, an increase in efficiency and areal density capability (ADC) is needed to keep up with this demand. The ADC of HDDs is achieved through increasing the bit and track densities, which requires the media grain size reduction, and increase of magnetocrystalline anisototropy to prevent thermal instabilities due to superparamagnetic effects \cite{FePt_HAMR,Magnetic_Prop}. The ‘magnetic recording trilemma’ describes the  conflicting requirements for recording density, thermal stability and writeability \cite{Pulsed_HAMR}. The requirement of small grain size necessitates an increase in the grain's magnetocrystalline anisotropy, in turn requiring a stronger magnetic field for the grain's reversal, which exceeds that achievable with integrated electromagnetic perpendicular magnetic recording (PMR) write heads. One possible solution to this, pioneered by Seagate, is using heat assisted magnetic recording (HAMR) \cite{HAMR_review}. The magnetic track is heated locally by a laser to a temperature above its Curie temperature $T_C$, thus temporarily removing anisotropy to facilitate the grains' switching within the heated spot \cite{HAMR_review}. A common recording material used in HAMR is FePt due to the optimal combination of high anisotropy, reasonably good spontaneous magnetization and low Curie temperature of the order of 700~K \cite{FePt_HAMR}. 
\\

Standard micromagnetic modeling is based on numerical solution of Landau-Lifshitz-Gilbert (LLG) equation in the assumption of constant magnetization (macrospin) of a unit cell used for meshing the ferromagnet \cite{Fidler_2000}. Micromagnetic modeling of HAMR recording processes is significantly more complicated due to the recording happening at around $T_C$ and above, where the assumption of constant spontaneous magnetization $M_s$ is not valid. A number of approaches have been used to address this challenge, including Landau-Lifshitz-Bloch (LLB) equation enabling the temperature dependent magnetization vector length \cite{LLB_Garanin, LLB_Chubykalo, LLB_HAMR, LLB_HAMR_Brill} or LLG with renormalized parameters (magnetization, anisotropy, exchange stiffness, damping) at elevated temperatures \cite{rLLG, Vitora_HAMR}. Both these approaches require a number of assumptions for HAMR modeling, most importantly, that of some finite $M_s$ value above $T_C$, which is not physical.
\\

Another approach to modeling of magnetic materials is atomistic spin dynamics (ASD), which implies localisation of electrons, confining the magnetic moment in the crystal lattice sites \cite{Ellis_ASD}. This is a perfectly valid assumption for local ferromagnets, but it was demonstrated that itinerant 3d magnetic materials are also well described by this phenomenology \cite{ASD_Evans}. An atomistic magnetic moment is always constant in ASD, and the materials' temperature dependent  parameters, including the ordering temperature $T_C$, naturally emerge due to accurate inclusion of stochasticity.
\\

An important use of ASD is to understand the effects of finite size on the properties of FePt. In particular, the distribution of $T_C$ has been investigated, and it was found that a rapid decrease in $T_C$ is obtained as the grain size of FePt becomes smaller \cite{Tc_FePt_grains}. ASD modeling has also been used to discover other important factors for HAMR such as the damping mechanism which was found to rapidly increase as it approaches $T_C$ and is further enhanced by reduction in grain size \cite{Damp_Anis_FePt}. This makes using ASD for HAMR modeling even more beneficial, since it allows accurate modeling of thermal gradients and transition jitter effects, which depend on the grains' $T_C$ and anisotropy distribution, and which determine media signal-to-noise ratio, bit error rate (BER), and other important recording parameters \cite{ThermGrad_HAMR,TransCurv_HAMR,SNR_HAMR}. Another important advantage of the atomistic discretization is the more realistic description of the structure and shape of small grains and their boundaries in high ADC media. The biggest disadvantage of atomistic modeling is the required computational power and time, making purely atomistic models not usable for recording real size tracks. Limited multiscale usage of HAMR processes have been previously explored \cite{Kazantseva_FePt,Meo_HAMR}, where an atomistic model was used to parameterize a micromagnetic model.
\\

In our work, we propose a true multiscale model, where simultaneous atomistic and micromagnetic modeling regions are used inside a moving simulation window, embedded within a granular magnetic track. This approach excels since the atomistic region is reserved for high temperatures around $T_C$, whereas the micromagnetic regions with temperature-dependant magnetic parameters are reserved for the lower temperature regions. This multiscale HAMR track writing algorithm is developed to demonstrate the bit writing process required for data storage, implemented in the multiscale micromagnetic/atomistic modeling software BORIS \cite{BORIS} using graphical processing units (GPU). This approach is favourable as it utilises crucial aspects of each modeling type whilst reducing the drawbacks, such as excessive computation time and inaccuracy.

\section{HAMR Multiscale Algorithm}
\begin{figure*}[h!]
	\includegraphics[width=1.0\textwidth]{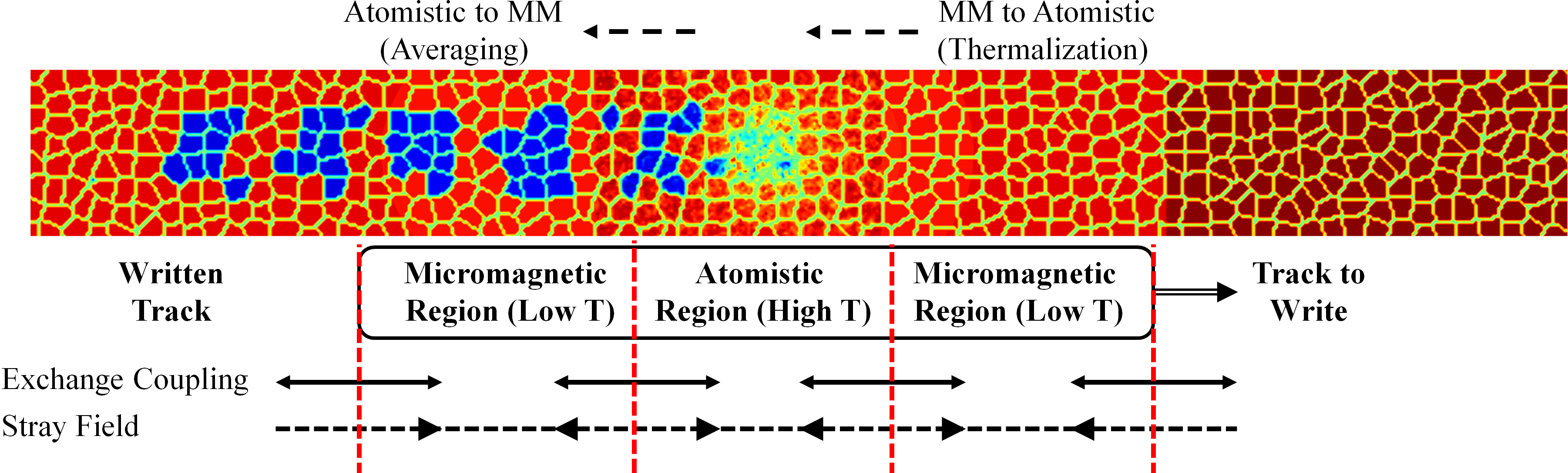}
	\caption{Diagram of multiscale HAMR modeling track writing algorithm, with a simulation region embedded within a granular magnetic track. The simulation region consists of a central atomistic modeling region for higher temperatures, with left and right micromagnetic modeling regions for lower temperatures, with region boundaries indicated by vertical dashed lines. All regions are exchange-coupled across the boundaries and with stray field contributions fully computed. The simulation region moves along the track at a set velocity, with magnetization from the micromagnetic region transferred to magnetic moments in the atomistic region using a thermalization process, and magnetic moments from the atomistic region transferred to magnetization in the micromagnetic region using averaging.}\label{Fig:algorithm}
\end{figure*}

HAMR modeling  is implemented in the following way – a high temperature spot (either introduced inside the model or imported from an external simulation) overlapping a writer's magnetic field profile is moved along a granular magnetic track of exchange-decoupled FePt grains, obtained by Voronoi construction, with a set velocity, and a bit sequence written by changing the magnetic field polarity. This may be done with a micromagnetic formalism, or using ASD modeling. The problem with either approach is the size of the simulation mesh required for writing long bit sequences,  as needed to determine BER \cite{BER} for example, which results in unacceptably long simulation times. Moreover, if thermally activated magnetic switching events, and distributions of temperature-dependent magnetic parameters are to be included accurately, particularly as the temperature approaches and exceeds $T_C$, the ASD formalism is preferable. However, ASD simulations can be over an order of magnitude slower than micromagnetic simulations. Here we solve both these problems by introducing a multiscale HAMR track writing algorithm, as depicted in Figure \ref{Fig:algorithm}.
\\

A simulation window is defined, which is embedded in a long magnetic track. The temperature and external magnetic field profiles are centred in the simulation window, which moves along the track at a set velocity in steps set by the track discretization cellsize. The magnetic track is effectively frozen, thus the computational cost is set by the size of the simulation window alone, and not by the length of track being written. As the simulation window moves, magnetization is transferred from the track to the window at the leading end, and written magnetic information is transferred from the simulation window to the track at the trailing end. Whilst the magnetic track is frozen, the stray field generated by it is fully computed when the simulation window is shifted, and included in the simulation window at each iteration. Moreover, the ends of the simulation window are exchange coupled to the track (further details below). The simulation window can consist of a single computational mesh – either fully micromagnetic or fully atomistic – or multiple meshes. Thus, the simulation window itself can be multiscale, and an effective configuration here consists of a central atomistic mesh which contains the high temperature region, with micromagnetic meshes either side for the lower temperature regions, exchange coupled to the atomistic mesh, and with the stray field between meshes fully computed each iteration. The atomistic mesh can be defined with any crystal structure required, and here we investigate the simple cubic (SC) structure, as well as the realistic L\textsubscript{10} structure for FePt \cite{MProp_FePt_HAMR}, with unit cell detailed in Appendix A.  Inside the atomistic mesh the magnetic spins are evolved using the stochastic LLG equation \cite{sLLG}:

\begin{equation}
\frac{\partial \mathbf{m}}{\partial t} = -\gamma \mathbf{m} \times (\mathbf{H} + \mathbf{H}_{th}) + \lambda \mathbf{m} \times \frac{\partial \mathbf{m}}{\partial t}
\label{eq:1}
\end{equation}

Here \textbf{m} is the magnetic spin direction, $\lambda$ is the damping, and $\gamma$ is the gyromagnetic ratio. The thermal stochastic field, \textbf{H}\textsubscript{th}, is generated using a normal distribution with zero mean and standard deviation $\sqrt{2 \lambda k_B T / \gamma \mu _0 \mu _S \Delta t}$, where $\mu _S$  is the magnetic moment, $T$ is the temperature and $\Delta t$ is the integration time-step. In the micromagnetic regions either the micromagnetic LLG equation \cite{rLLG} – same form as Equation (1), but without the stochastic field – or the Landau Lifshitz Bloch (LLB) \cite{LLB_Garanin} is solved, which additionally contains a longitudinal relaxation term, and here we investigate both. In the micromagnetic regions it is further necessary to impose temperature dependences of magnetic parameters. These are $M_S(T) = M_S^0 m_e(T)$ for the spontaneous magnetization, $A(T) = A_0 m_e^2(T)$ for the exchange stiffness \cite{A_tdep}, $K(T) = K_0 m_e^3(T)$ for the uniaxial anisotropy constant \cite{CallenCallen}, and $\alpha = \lambda(1-T/3T_C)$ for the micromagnetic Gilbert damping with $T < T_C$. Note, the micromagnetic damping term coincides with the atomistic damping term at 0 K, with the latter being independent of temperature. Here $m_e(T) = B[m_e 3 T_C / T + \mu _S \mu _0 H_{ext} / k _B T]$ is the equilibrium magnetization scaling function, where $T_C$ is the Curie temperature, $H_{ext}$ is the external magnetic field magnitude, and $B(x) = coth(x) - 1/x$ is the Langevin function. An advantage of ASD modelling is the natural emergence of magnetic parameters temperature dependences, whilst for micromagnetic modelling these must be enforced using approximate relations as given above, and thus use of an atomistic modelling window in the multiscale algorithm leads to more accurate results. Another option for micromagnetic modelling is computation of temperature dependences using atomistic modelling \cite{Kazantseva_FePt,Meo_HAMR}, or use of empirical data to introduce temperature dependences.
\\

In the atomistic simulation region the effective field \textbf{H} in Equation \eqref{eq:1} contains a number of contributions, including direct exchange, uniaxial anisotropy, stray field from the track and micromagnetic regions, demagnetizing field (here approximately the same as the field due to dipole-dipole interaction), as well as the external write field. The direct exchange field at a spin \textit{i} is given in Equation \eqref{eq:2}, where the sum runs over the neighbours $j \in N$, and $J = J_{ij}$ is the exchange energy. For the SC structure the 6 nearest neighbours are considered. For the L\textsubscript{10} FePt structure the 4 nearest neighbours, as well as the 6 next nearest neighbours are included.

\begin{equation}
\mathbf{H}_{exch,i} = \frac{1}{\mu _0 \mu _S} \sum_{j \in N} J_{ij} \mathbf{m} _j
\label{eq:2}
\end{equation}

The uniaxial anisotropy interaction field is given in Equation \eqref{eq:3}, where $K_1$ is the anisotropy energy and $\mathbf{e}_A = \hat{\mathbf{z}}$ is the perpendicular anisotropy axis direction.

\begin{equation}
\mathbf{H}_{an,i} = \frac{2 K_1}{\mu _0 \mu _S} (\mathbf{m} _i \cdot \mathbf{e}_A)\mathbf{e}_A
\label{eq:3}
\end{equation}

In the micromagnetic region these interactions become the micromagnetic exchange interaction – Equation \eqref{eq:4} – and the micromagnetic uniaxial anisotropy interaction – Equation \eqref{eq:5}.

\begin{equation}
\mathbf{H} = \frac{2 A}{\mu _0 M_S} \nabla^2 \mathbf{m}
\label{eq:4}
\end{equation}

\begin{equation}
\mathbf{H} = \frac{2 K}{\mu _0 M_S} (\mathbf{m} \cdot \mathbf{e}_A)\mathbf{e}_A
\label{eq:5}
\end{equation}

For exchange coupling the atomistic and micromagnetic regions Equations \eqref{eq:2} and \eqref{eq:4} are used, respectively. Equation \eqref{eq:4} is evaluated in a finite-difference implementation with a 6-neighbor stencil for the Laplacian operator. For the contribution from the atomistic region the average atomistic spin direction is obtained in the respective stencil cell. For the contribution from the micromagnetic region the respective magnetization direction is used directly since the atomistic lattice constant is smaller than the micromagnetic cellsize.
\\

When transferring from an atomistic to a micromagnetic region the micromagnetic magnetization vector $\mathbf{M}$ is obtained as the sum of atomistic magnetic moments divided by the micromagnetic cellsize volume $V$, as shown in Equation \eqref{eq:6}.

\begin{equation}
\mathbf{M} = \frac{\mu _S}{V} \sum_{j \in V} \mathbf{m} _j
\label{eq:6}
\end{equation}

When transferring from a micromagnetic to an atomistic region, the same Equation \eqref{eq:6} is used, but now as an inverse problem. The atomistic spins are generated using a uniform distribution of the polar angle from the magnetization vector direction, inside a cone with a set maximum angle. The maximum angle required to reproduce the correct micromagnetic magnetization magnitude \textit{M} is obtained as $\theta _{max} = (5 \pi / 9) \sqrt{1 - \sqrt{1 + 6(M/M _S ^0 - 1) / 5}}$, where $M_{S}^{0}$ is the magnetization at zero temperature, obtained in the ground state when all the atomistic spins are aligned.
\\

For micromagnetic regions the demagnetizing field is computed using Equation \eqref{eq:7}, using the standard approach of a convolution sum with the demagnetizing tensor $\mathbf{N}$ \cite{Newell}. 

\begin{equation}
\mathbf{H}_{d,i} =- \sum _{j} \mathbf{N}(\mathbf{r}_j - \mathbf{r}_i) \mathbf{M} _j
\label{eq:7}
\end{equation}

For atomistic regions, rather than using the dipole-dipole interaction, the same Equation \eqref{eq:7} is used, where the atomistic magnetic moments are first transferred to the coarser micromagnetic discretisation using Equation \eqref{eq:6}. As we have verified the demagnetizing field obtained using Equation \eqref{eq:7} is approximately the same as that obtained using the more expensive dipole-dipole interaction. This approach also allows computation of the stray field between multiple micromagnetic and atomistic meshes, which is done using the multilayered convolution algorithm previously introduced \cite{multiconv, multiGPU}. Thus, within the simulation region the demagnetizing and stray fields are re-computed at every iteration, whilst the stray field contribution from the track is updated only when the simulation region moves. The algorithm has been implemented for central processing units, single GPUs, as well as multiple GPUs \cite{multiGPU}.

\section{HAMR Simulations}

Employing atomistic modeling in HAMR simulations could reveal previously hidden interactions between the heating of the medium and magnetization reversal. Therefore, multiscale modeling is a favourable approach since the temperatures will be extreme at the centre of the laser spot, whereas the outer areas will have a lower temperature which is adequately described by micromagnetic modeling. This section outlines the results obtained from these various types of modeling.
    
\begin{figure}[h!]
	\includegraphics[width=0.5\textwidth]{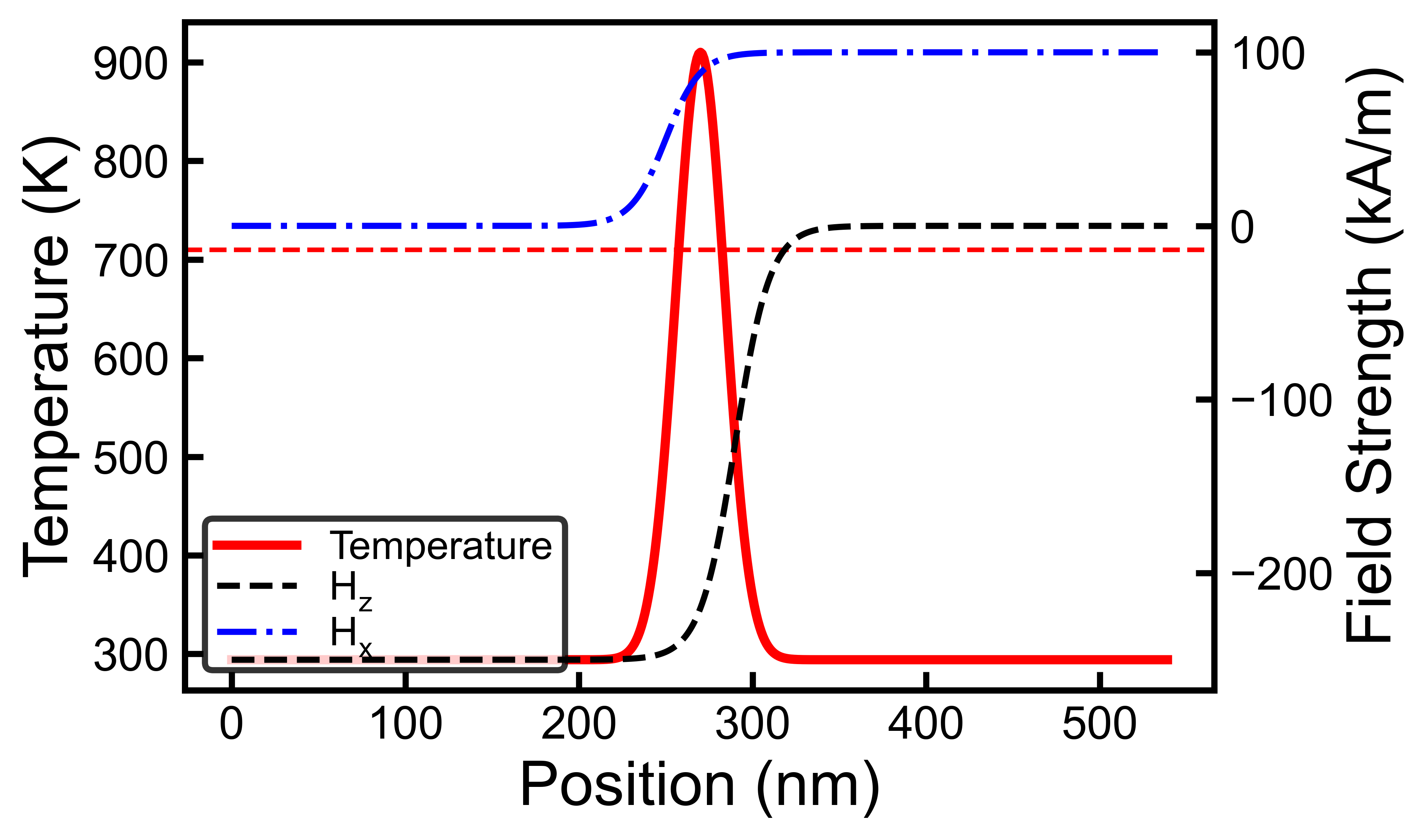}
	\caption{Plot of the Gaussian temperature profile due to the laser spot (solid red line), shown along the track, with a horizontal dashed red line to indicate the Curie temperature at 710 K. The black dashed line shows the out-of-plane component ($H_z$) of the field profile along the track, whilst the dash-dot blue line shows the in-plane component ($H_x$).}\label{Fig:profiles}
\end{figure}

The HAMR simulations for this paper were performed using idealized temperature and field profiles, plotted in Figure \ref{Fig:profiles}. This figure shows the Gaussian temperature profile due to a laser spot, used to heat the magnetic track above $T_C$. Figure \ref{Fig:profiles} has a horizontal marker (dotted red line) to indicate $T_C$ at 710K, and the laser spot diameter above $T_C$ was 25 nm, which largely determined the width of written bits. The magnetic field from the write pole was simulated as $tanh$ profiles along the track, both for the out-of-plane and in-plane field components shown in Figure \ref{Fig:profiles}. As the shape and positioning suggest, the field is used to reverse the grains which have been heated. The field strength is not required to be large due to the decrease in coercivity from heating, so unheated grains are not reversed, with spatial resolution provided by the temperature spot. 
\\

The magnetic tracks used throughout the simulations had dimensions of 540 nm $\times$ 60 nm $\times$ 16 nm with cell sizes of 1 nm $\times$ 1 nm $\times$ 2 nm, and a grain size of 8 nm, with a simulation window length of 300 nm. Arbitrarily long tracks may be set, without increasing the time per iteration. However, this track length was sufficient to accommodate a varied bit pattern. Atomistic simulations had a SC structure with lattice constant $a$ = 0.5 nm. Other grain sizes of 7 nm and 9 nm were also simulated. The laser spot moved along the track at a velocity of 20 m/s, comparable to the linear velocity in a physical HDD.
\\

For the atomistic models, the exchange energy for the spin moments was obtained by using the desired $T_C$ of 710 K in conjunction with Equation \eqref{eq:8}. However, the mean-field correction factor ($\epsilon$) is unknown, so to obtain a value for this and thus the exchange energy, Monte Carlo modeling was used \cite{Lepadatu_MC}. The mean-field correction factor is dependent on the crystal structure. The resulting mean-field correction factors were determined to be $\epsilon _{SC}$ = 0.72 and $\epsilon _{L10}$ = 0.78 for SC and L$_{10}$ respectively, obtaining exchange energies $J_{SC} = 6.8 \times 10^{-21} J$ and $J_{L10} = 3.77 \times 10^{-21} J$. The co-ordination number ($z$) for SC is 6 and 10 for L\textsubscript{10}. In this work, the exchange energy was set to be identical for all included neighbours. However, further work could use a more advanced model to take into account the varying $J$ values of the nearest and further distant neighbours. \cite{Evans_PhD}

\begin{equation}
J = \frac{3 k_B T_C}{\epsilon z}
\label{eq:8}
\end{equation}

For micromagnetic simulations, the exchange energy is related to exchange stiffness $A$, as shown in Equations \eqref{eq:9} and \eqref{eq:10} for SC and L\textsubscript{10} respectively. As shown in Appendix A, we have verified using spin-wave dispersion simulations that the implemented exchange interactions for SC and L\textsubscript{10} crystal structures are correct.

\begin{equation}
A_{SC} = \frac{J}{2 a}
\label{eq:9}
\end{equation}

\begin{equation}
A_{L10} = \frac{3J}{2 a}
\label{eq:10}
\end{equation}

The $M_S$ and $K$ micromagnetic parameters were used as $5.2 \times 10^5 A/m$ and $2.2 \times 10^6 J/m^3$ respectively. These micromagnetic parameters also have a temperature dependance as given in the previous section. A damping value of 0.1 (at 0 K) was used for both atomistic and micromagnetic regions, with the latter also having a temperature dependence as described above.
\\

\begin{figure*}[h!]
	\includegraphics[width=1.0\textwidth]{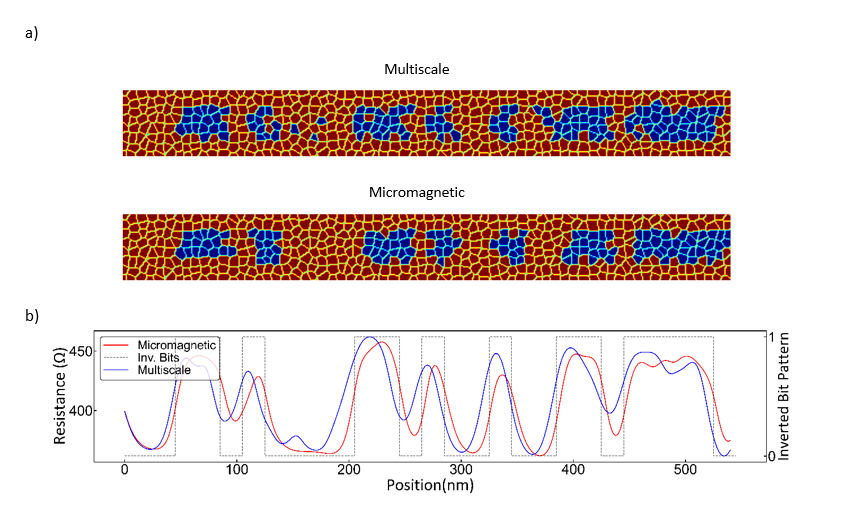}
	\caption{a) Written bit patterns on a granular track with 8 nm grain size, for a uniform out-of-plane magnetized initial state (red), shown for multiscale and micromagnetic LLG simulations. The into-the-plane grains (blue) represent reversed grains. b) Plots of a TMR read-head scan obtained for the respective bit patterns shown in a), with the inverted bit pattern also plotted.}\label{Fig:sim_unif}
\end{figure*}

Figure \ref{Fig:sim_unif} shows the results of the multiscale algorithm described in Figure \ref{Fig:algorithm}. The atomistic central window used a SC crystal structure while the windows on either side were micromagnetic LLG. The grain sizes were not uniform along the track, with many being larger or smaller than the nominal value of 8 nm. This causes a distribution in temperature dependence along the track which means that there is a distribution in $T_C$. The micromagnetic track in Figure \ref{Fig:sim_unif} has been simulated with a $T_C$ distribution, in particular a normal distribution with standard deviation about $T_C$ = 710 K of $0.05 \times T_C$. For comparison purposes the bit pattern to be written was kept consistent throughout all simulations. The bit pattern used was 0010111100101101100100001 and the bit length was 20 nm. The red grains represent the out-of-plane direction, the blue grains represent the reversed into-the-plane direction, and the green areas represent empty space. Whilst the simulated track width was 40 nm, the written bits were approximately 25 nm wide, which reflects the diameter of the laser spot above $T_C$.
\\

A simulated TMR read-head was used to read the information stored on the track. This read-head traversed the tracks along its length and converted the stray field due to the out-of-plane magnetization to a resistance value. The TMR read-head consists of a magnetic tunnel junction sensor on top of a synthetic antiferromagnetic (SAF) stack and with permanent magnet shields to protect against unwanted stray fields, as detailed previously \cite{BORIS_MTJ}. These data sets are plotted in Figure \ref{Fig:sim_unif}(b). This plot shows that the overall pattern is in line with the expected bit pattern and is similar to the micromagnetic result.
\\

The multiscale method excels in these circumstances, as at high temperatures atomistic modeling is important due to stochasticity. From the visual comparison, the simulations resulted in tracks with a clear comparable magnetic reversal. The pre-defined bit pattern is consistent through both images and follows the expected sequence. However, there are two consistent differences. As observed in Figure \ref{Fig:sim_unif}(b) the reader pattern obtained from the micromagnetic HAMR write consistently lags behind that obtained from the multiscale HAMR write. This is due to different threshold temperatures required to obtain grain switching in atomistic modelling, compared to the less accurate micromagnetic model with enforced magnetic parameter temperature dependences. In particular, with ASD grain switching is obtained at lower temperatures compared to the micromagnetic model, resulting in approximately 1 grain shift to the left. Moreover, the multiscale (atomistic) track is less ordered than its micromagnetic counterpart, showing random grain switching. This is due to thermally activated switching, which describes random flipping of spins at higher temperatures related to superparamagnetic phenomena \cite{Magnetic_Prop}. This is further evidenced by simulations with 7 nm and 9 nm grain sizes. These results show that the larger the grain size, the fewer thermally activated switched grains are present. Such thermally activated switching occurs only in the vicinity of $T_C$ for ASD/multiscale modelling owing to stochasticity introduced in the LLG equation. For comparison purposes, micromagnetic simulations were run using LLG and LLB equations. Stochastic micromagnetic modeling is also possible. However, this is beyond the scope of the current work. Both equations induce switching using the transverse damping term but the main difference in these approaches is the existence of a longitudinal relaxation term in LLB, which is particularly important in the ultrafast time regime. In the simulations, the parameters changed between methods was the time-step used to solve the equations using the RK4 method. LLB required a significantly smaller timestep of 2.5 fs whereas LLG needed a 20 fs timestep. This is reflected in the time taken for simulations to complete, which were run on a single A5000 GPU. LLG, with the larger time-step, completed the simulation in approximately 40 minutes whereas LLB took 440 minutes. The resulting simulations showed that the different equations do not differ significantly in clarity of writing bits. Considering the significant additional time taken for LLB simulations and the dominant precessional switching at this timescale, we proceeded to use LLG for the micromagnetic regions in the multiscale algorithm.
\\

In the same manner, atomistic crystal structures of SC and L$_{10}$ were compared to investigate the advantages and disadvantages of using realistic FePt L\textsubscript{10}. L\textsubscript{10} required a smaller time-step of 5 fs as the solution diverged when using the same time-step of SC at 10 fs. The time taken for completion of SC simulation was approximately 295 minutes and 1686 minutes for L\textsubscript{10}. The simulations resulted in similar tracks with minor differences which can be accounted for by thermally activated switching. In terms of efficiency, SC is the reasonable crystal structure to use for multiscale simulations. However, further work is needed to investigate the effect of crystal structure and exchange constants parameterization in Equation \eqref{eq:2}. An important difference between SC and L\textsubscript{10} is the longitudinal magnetic susceptibility, which, as shown in Appendix A, is around one order of magnitude smaller for L\textsubscript{10}. This may be important if ultrafast processes are considered, however in the present case where precessional switching dominates, the results obtained show the SC structure is sufficient. For the multiscale modeling algorithm, the 300 nm long simulation window is split between a 100 nm long atomistic simulation region and two 100 nm long micromagnetic simulation regions. With an SC structure, the simulation time was 228 minutes, which is a modest decrease compared to a fully SC atomistic window. However, the performance is strongly dependent on the size and complexity of the problem. For example, with the more complex L\textsubscript{10} atomistic region, the multiscale algorithm completed in 459 minutes, which is between three to four times faster compared to a fully L\textsubscript{10} atomistic window. The algorithm can be further sped up using multiple GPUs, which is particularly beneficial for large simulation windows \cite{multiGPU}. Even for the current relatively small simulation windows, a further speedup factor up to 40\% is achieved with 2 $\times$ A5000 GPUs.
\\

\begin{figure*}[h!]
	\includegraphics[width=1.0\textwidth]{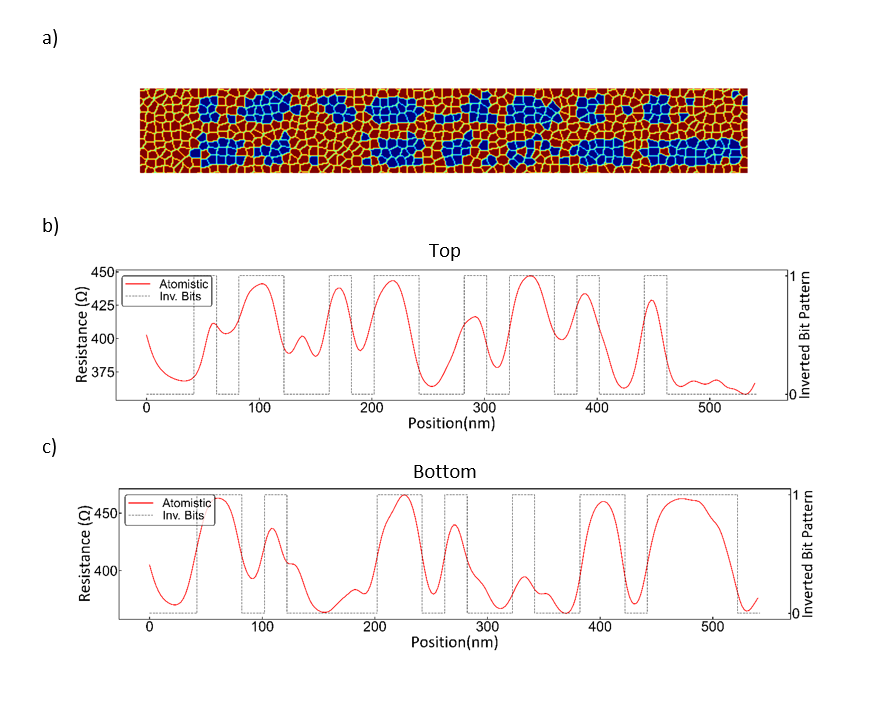}
	\caption{a) Written bit patterns with two tracks. The into-the-plane grains (blue) represent reversed grains. Plots of a TMR read-head scan obtained for the respective bit patterns shown in a), with the inverted bit pattern also plotted for b) top track, and c) bottom track.}\label{Fig:sim_multi}
\end{figure*}

Finally, the multiscale track shifting algorithm is applied for writing parallel tracks, with results shown in Figure \ref{Fig:sim_multi} for two parallel tracks. Here, the bits are approximately 25 nm wide as before, with the separation between tracks being 15 nm, for a total simulation window with 80 nm width. The two tracks were written sequentially. Starting from the uniform magnetization state, the bottom track was written first as before, achieved by positioning the laser spot in the centre of the bottom track. The written bit pattern was then used as the starting state for a second pass through the track shifting algorithm, this time writing a different bit pattern in the top track. The respective bit patterns are shown in Figure \ref{Fig:sim_multi}(b) and Figure \ref{Fig:sim_multi}(c) for the top and bottom tracks respectively. The TMR read-head scans show that the written bit patterns were reproduced, with transitions between bits clearly identified. Thus, the multiscale track shifting algorithm developed here can also be used to study the effect of writing multiple parallel tracks. Whilst the 15 nm separation between tracks does not affect previously written bit patterns, with the read-head scan resolution also preserved, further studies can be conducted to investigate the effect of reducing track-track separation, or even shingled magnetic recording, in order to identify acceptable operating parameters as the areal density is increased.

\section{Conclusion}
In summary, through this work, the concept of using a multiscale algorithm combining atomistic and micromagnetic modeling has been demonstrated by simulating the bit writing process in HAMR on a simulated FePt track, both on single tracks as well as parallel tracks. The simulations were quantitatively analysed through a simulated TMR read-head. An atomistic modeling region was used for higher temperatures around the laser spot, whilst micromagnetic regions were used for lower temperatures either side of the atomistic region. This multiscale approach is favourable due to the additional accuracy compared to purely micromagnetic simulations, as atomistic modeling at higher temperatures naturally includes the effects of stochasticity, suitable both for the ferromagnetic and paramagnetic phases below and above the Curie temperature respectively. Atomistic modeling at higher temperatures accurately includes the effect of thermally activated grain switching during the HAMR bit writing process, which was found to be dependent on the grain size. Moreover, the general purpose multiscale HAMR track writing algorithm developed here, and implemented in the micromagnetic/atomistic multiscale software BORIS, allows the atomistic modeling region to be configured with a realistic crystal structure, depending on the material investigated, as well as any configuration of exchange interactions between neighbors. Here, we have investigated the L\textsubscript{10} FePt crystal structure, and compared it to the simpler SC structure, and whilst the written bit patterns did not differ between the two approaches, further studies using the multiscale algorithm developed here could benefit from inclusion of advanced atomistic modeling regions. On the other hand, inclusion of micromagnetic modeling regions allows a larger simulation window to be defined, with reduced computation time compared to a purely atomistic modeling approach.

\section{Appendix A}

\begin{figure}[h!]
	\includegraphics[width=0.5\textwidth]{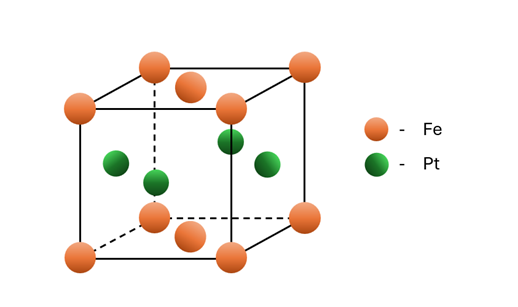}
	\caption{Schematic diagram of a L$_{10}$-ordered FePt crystal lattice.}\label{Fig:crystal}
\end{figure}

For atomistic modeling two crystal structures were considered: simple cubic (SC) and L$_{10}$-ordered FePt – the latter is shown in Figure \ref{Fig:crystal}. Here we only model the Fe atomic sites, with their magnetic moments set such that $M_S = 520 \times 10^5 A/m$ is obtained. For SC this means $\mu _S = M_S a^3$, whereas for FePt L$_{10}$ we have $\mu _S = M_S a^3/2$, since each unit cell contains 2 Fe atoms. Similarly, for the micromagnetic uniaxial anisotropy energy density $K = 2.2 \times 10^6 J/m^3$, we have $K_1 = K a^3$ for the SC anisotropy energy, and $K_1 = K a^3/2$ for FePt L$_{10}$.
\\

\begin{figure}[h!]
	\includegraphics[width=0.5\textwidth]{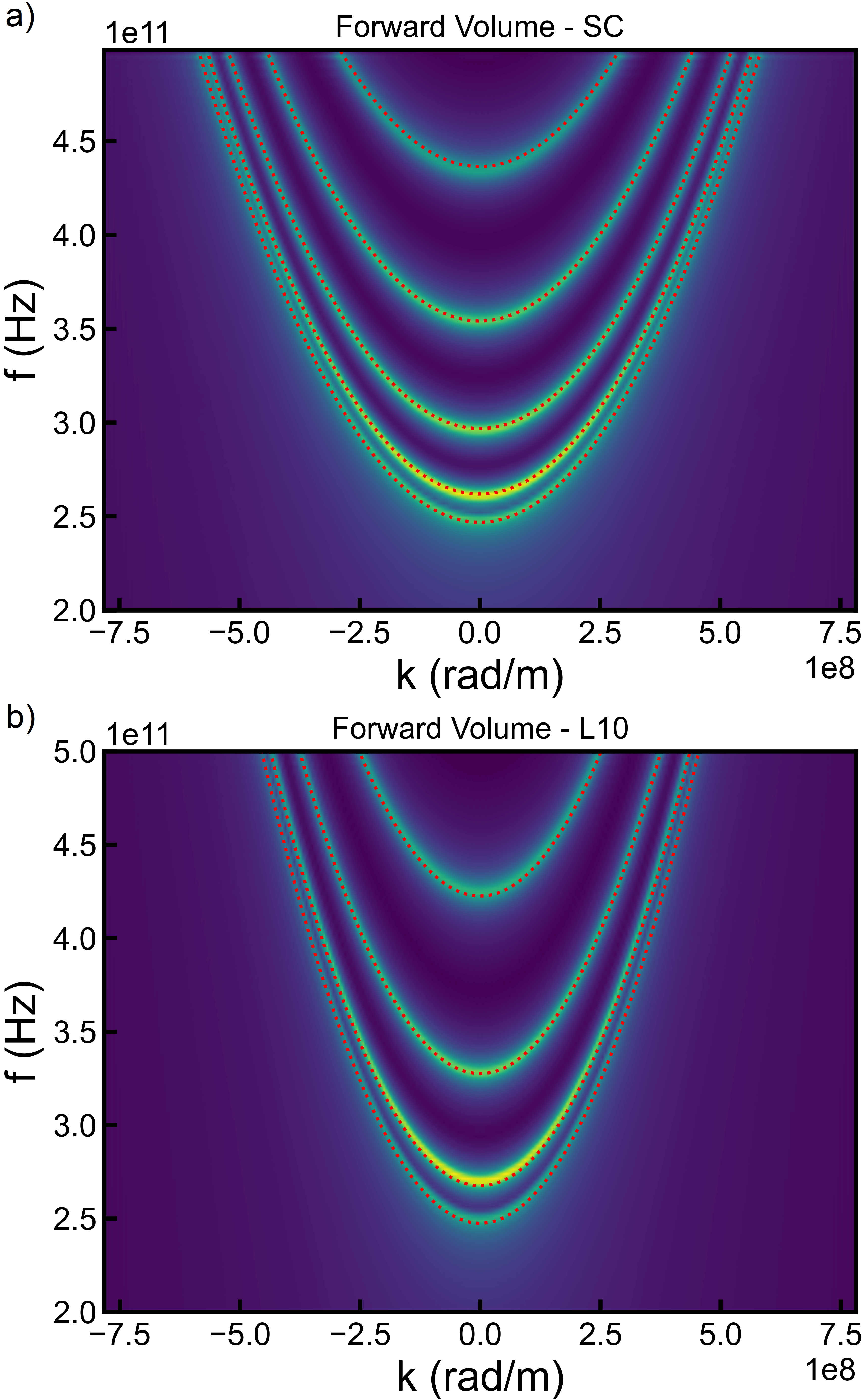}
	\caption{Heat map displaying the obtained dispersion curves from spin-wave excitations in the forward volume for SC and L$_{10}$ crystal structures.}\label{Fig:SW}
\end{figure}

To ensure the exchange coupling and exchange stiffness values used are correct, we simulated spin-wave dispersions to corroborate simulation results to theoretical expectations. The methodology used is outlined in \cite{SWdispersion}. This was done in BORIS by using a thin magnetic material with dimensions of 1000 nm x 50 nm x 1 nm, on which a $sinc$ pulse was used to excite spin-waves in the frequency domain in the forward volume (FV) geometry. The FV geometry describes the wavevector along the longitudinal direction and the bias field in the out-of-plane direction, perpendicular to the wavevector. The spin-wave dispersions are parabolic in the frequency–wavevector ($f-k$) space as described by Equation \eqref{eq:11}. These equations are important as they outline the relationship needed to verify the exchange values used. The exchange factor, $\lambda _{ex}$, depends on the exchange stiffness, $A$, which is  directly proportional to the exchange coupling. A further consequence is the dependance on crystal structure due to varying lattice geometry shown in Equations \eqref{eq:9} and \eqref{eq:10} for SC and L$_{10}$ respectively. Figure \ref{Fig:SW} displays the obtained heat map for the dispersion curves of the two crystal structures with the theoretical curves superimposed as red dots. The results show that theoretical and simulated curves align well, verifying the implemented exchange interaction.
\\

\begin{equation}
\begin{split}
& \omega _{ex} = \omega _0 + \lambda _{ex} \gamma M_S k^2 ,\\
& \lambda_{ex} = 2A / \mu _0 M_s^2
\end{split}
\label{eq:11}
\end{equation}

\begin{figure}[h!]
	\includegraphics[width=0.5\textwidth]{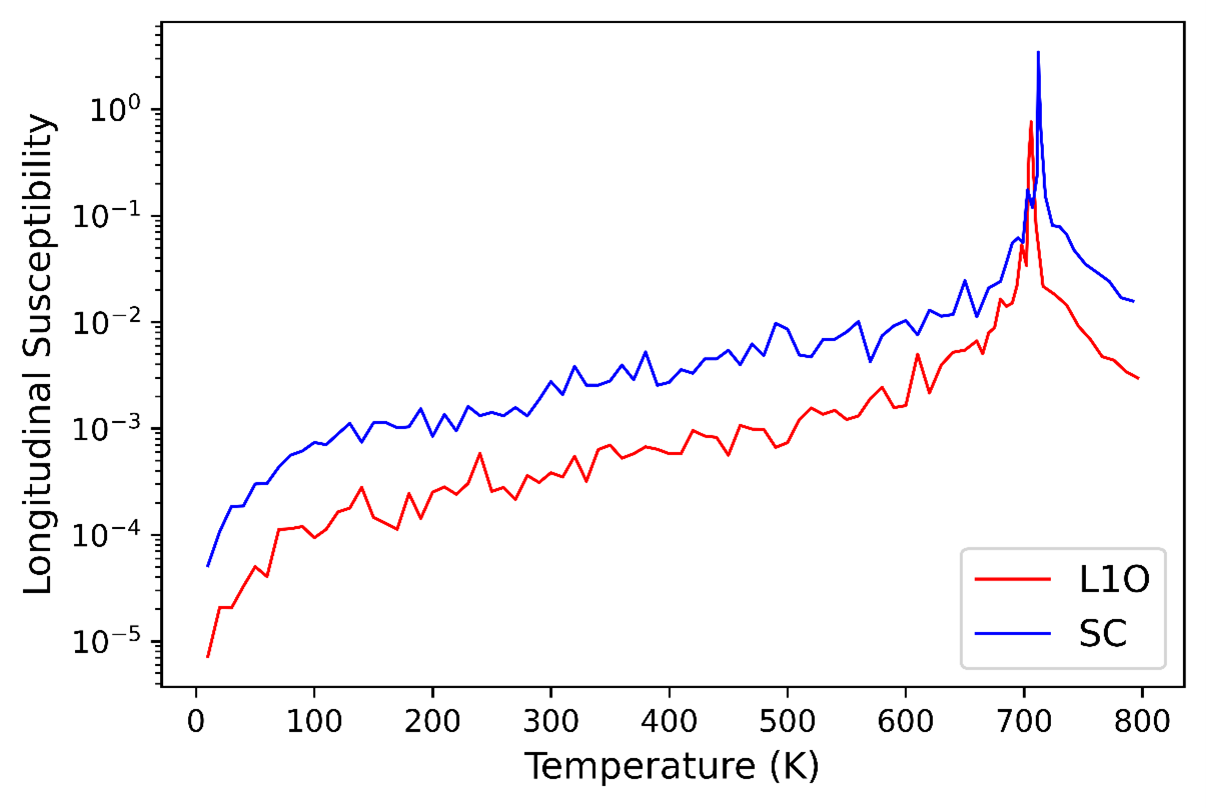}
	\caption{Longitudinal magnetic susceptibility of SC and L$_{10}$ ordered crystal structures, plotted on a logarithmic scale.}\label{Fig:Susceptibility}
\end{figure}

The longitudinal magnetic susceptibility is an important property to investigate when looking at a magnetic material. This property determines how sensitive the magnetization length is to external applied fields such as the one used in HAMR, therefore, in this work we modelled the longitudinal magnetic susceptibility of L$_{10}$ and SC crystal structures. This was done using a Monte Carlo algorithm implemented in BORIS, as described previously \cite{Lepadatu_MC}, with results shown in Figure \ref{Fig:Susceptibility}. It was found that there is an order of magnitude difference between the susceptibilities of L$_{10}$ and SC, which is particularly important in the ultrafast regime where longitudinal magnetization switching processes are important. However, for HAMR modeling where precessional switching dominates, it was determined that the use of realistic crystal structure has a negligible effect on the outcome of the written bit pattern. For this reason the SC crystal structure is preferable for HAMR modeling due to significantly decreased computation time.

\end{document}